\documentclass[aps,prl,longbibliography,superscriptaddress,reprint]{revtex4-2}

\usepackage{graphicx}
\usepackage{amsmath}
\usepackage{multirow} 
\usepackage{physics} 
\usepackage{subfigure}   
\usepackage{floatrow}
\usepackage{verbatim}
\usepackage{bm}
\usepackage{soul}
\usepackage{color}
\usepackage[colorlinks,urlcolor=blue,citecolor=blue,linkcolor=blue]{hyperref}

\begin{document}

\title{Programmable simulation of high-order exceptional point with a trapped ion}

\author{Yue Li}\email{These authors contribute equally.}
\affiliation{CAS Key Laboratory of Microscale Magnetic Resonance and School of Physical Sciences, University of Science and Technology of China, Hefei 230026, China}
\affiliation{Anhui Province Key Laboratory of Scientific Instrument Development and Application, University of Science and Technology of China, Hefei 230026, China}

\author{Yang Wu}\email{These authors contribute equally.}
\affiliation{CAS Key Laboratory of Microscale Magnetic Resonance and School of Physical Sciences, University of Science and Technology of China, Hefei 230026, China}
\affiliation{Anhui Province Key Laboratory of Scientific Instrument Development and Application, University of Science and Technology of China, Hefei 230026, China}

\author{Yuqi Zhou}\email{These authors contribute equally.}
\affiliation{CAS Key Laboratory of Microscale Magnetic Resonance and School of Physical Sciences, University of Science and Technology of China, Hefei 230026, China}
\affiliation{Anhui Province Key Laboratory of Scientific Instrument Development and Application, University of Science and Technology of China, Hefei 230026, China}

\author{Mengxiang Zhang}
\affiliation{CAS Key Laboratory of Microscale Magnetic Resonance and School of Physical Sciences, University of Science and Technology of China, Hefei 230026, China}
\affiliation{Anhui Province Key Laboratory of Scientific Instrument Development and Application, University of Science and Technology of China, Hefei 230026, China}
\author{Xingyu Zhao}
\affiliation{CAS Key Laboratory of Microscale Magnetic Resonance and School of Physical Sciences, University of Science and Technology of China, Hefei 230026, China}
\affiliation{Anhui Province Key Laboratory of Scientific Instrument Development and Application, University of Science and Technology of China, Hefei 230026, China}
\affiliation{Hefei National Laboratory, University of Science and Technology of China, Hefei 230088, China}

\author{Yibo Yuan}
\affiliation{CAS Key Laboratory of Microscale Magnetic Resonance and School of Physical Sciences, University of Science and Technology of China, Hefei 230026, China}
\affiliation{Anhui Province Key Laboratory of Scientific Instrument Development and Application, University of Science and Technology of China, Hefei 230026, China}
\affiliation{Hefei National Laboratory, University of Science and Technology of China, Hefei 230088, China}

\author{Xu Cheng}
\affiliation{CAS Key Laboratory of Microscale Magnetic Resonance and School of Physical Sciences, University of Science and Technology of China, Hefei 230026, China}
\affiliation{Anhui Province Key Laboratory of Scientific Instrument Development and Application, University of Science and Technology of China, Hefei 230026, China}
\affiliation{Hefei National Laboratory, University of Science and Technology of China, Hefei 230088, China}

\author{Yi Li}
\affiliation{CAS Key Laboratory of Microscale Magnetic Resonance and School of Physical Sciences, University of Science and Technology of China, Hefei 230026, China}
\affiliation{Anhui Province Key Laboratory of Scientific Instrument Development and Application, University of Science and Technology of China, Hefei 230026, China}
\affiliation{Hefei National Laboratory, University of Science and Technology of China, Hefei 230088, China}

\author{Xi Qin}
\affiliation{CAS Key Laboratory of Microscale Magnetic Resonance and School of Physical Sciences, University of Science and Technology of China, Hefei 230026, China}
\affiliation{Anhui Province Key Laboratory of Scientific Instrument Development and Application, University of Science and Technology of China, Hefei 230026, China}
\affiliation{Hefei National Laboratory, University of Science and Technology of China, Hefei 230088, China}

\author{Xing Rong} \email{xrong@ustc.edu.cn}
\affiliation{CAS Key Laboratory of Microscale Magnetic Resonance and School of Physical Sciences, University of Science and Technology of China, Hefei 230026, China}
\affiliation{Anhui Province Key Laboratory of Scientific Instrument Development and Application, University of Science and Technology of China, Hefei 230026, China}
\affiliation{Hefei National Laboratory, University of Science and Technology of China, Hefei 230088, China}

\author{Yiheng Lin} \email{yiheng@ustc.edu.cn}
\affiliation{CAS Key Laboratory of Microscale Magnetic Resonance and School of Physical Sciences, University of Science and Technology of China, Hefei 230026, China}
\affiliation{Anhui Province Key Laboratory of Scientific Instrument Development and Application, University of Science and Technology of China, Hefei 230026, China}
\affiliation{Hefei National Laboratory, University of Science and Technology of China, Hefei 230088, China}

\author{Jiangfeng Du} 
\affiliation{CAS Key Laboratory of Microscale Magnetic Resonance and School of Physical Sciences, University of Science and Technology of China, Hefei 230026, China}
\affiliation{Anhui Province Key Laboratory of Scientific Instrument Development and Application, University of Science and Technology of China, Hefei 230026, China}
\affiliation{Hefei National Laboratory, University of Science and Technology of China, Hefei 230088, China}
\affiliation{Institute of Quantum Sensing and School of Physics, Zhejiang University, Hangzhou 310027, China}

\begin{abstract}
The nontrivial degeneracies in non-Hermitian systems, exceptional points (EPs), have attracted extensive attention due to intriguing phenomena. Compared with commonly observed second-order EPs, high-order EPs show rich physics due to their extended dimension and parameter space, ranging from the coalescence of EPs into higher order to potential applications in topological properties. However, 
these features also pose challenges in controlling multiple coherent and dissipative elements in a scaled system. Here we experimentally demonstrate a native programmable control to simulate a high-order non-Hermitian Hamiltonian in a multi-dimensional trapped ion system. We simulate a series of non-Hermitian systems with varied parameters and observe the coalescence of second-order EPs into a fourth-order EP. Our results pave the way for scalable quantum simulation of high-dimensional dissipative systems and can be beneficial for the application of high-order EPs in quantum sensing and quantum control.

\end{abstract}

\date{\today}

\maketitle  
Non-Hermitian physics has attracted widespread attention due to its potential applications such as unidirectional invisibility~\cite{Unidirect1, Unidirect2, Unidirect3, Unidirect4}, single-mode laser~\cite{Single_mode1}, wireless power transfer~\cite{Power_transfer1}, noise-assisted sensing~\cite{NA_sensing} and chiral heat transport~\cite{HeatTransport2023}. An exotic property of the non-Hermitian system is exceptional points (EPs) where both eigenvalues and eigenstates coalesce~\cite{RMP_EPtopology}. There have been investigations on new phenomena related to second-order EPs such as bulk Fermi arc~\cite{Fermic_arc1}, Weyl exceptional ring~\cite{Wely_ring1}, and topological phenomena when the EP is encircled~\cite{Brids1,encircle_EP1,encircle_EP2, encircle_EP3, encircle_EP4}. Various systems have been utilized to study second-order EP, such as nitrogen-vacancy center~\cite{NV_PT1}, electronic oscillators~\cite{Electronic_oscillator1},  ultracold atoms~\cite{Ultracold_atoms1, Ultracold_atoms2}, neutral atoms~\cite{Thermal_atoms} and trapped ions~\cite{Trapped_ions1, Trapped_ions2, Trapped_ions3, Trapped_ions4, Trapped_ions5}. 

\begin{figure}[t!]
	\begin{center}

	\includegraphics[width=1\columnwidth]{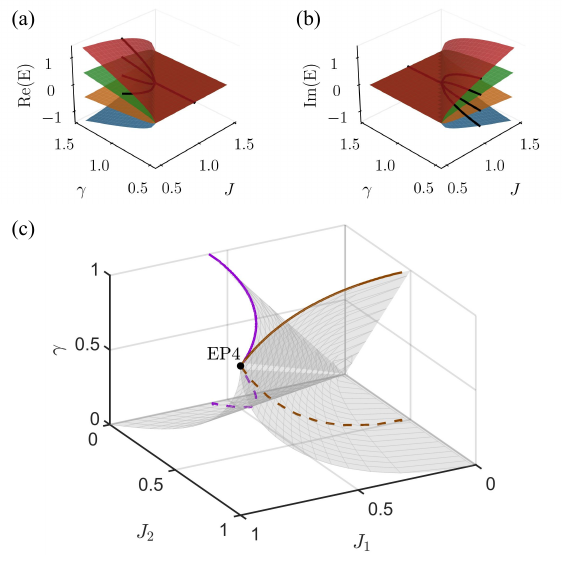}
	\caption{Band structure and EPs structure of the multilevel non-Hermitian Hamiltonian. (a)(b) Illustration of the real and imaginary components of the band structure with different coupling strength $J$ and dissipation rate $\gamma$. The EP4s locate at the line where $\abs{\gamma}=\abs{J}$. The black line represents the parameter trajectory demonstrated in this work. (c) The surfaces are composed of EP2s in the parameter space. Two types of lines composed of EP2s can be analytically solved when $\gamma=1$ and they would intersect at the point of EP4 (the purple solid line and the brown solid line in the plane of $\gamma=1$). These two lines projected along the negative gamma direction will outline two additional lines(dashed lines) on the surfaces of EP2s. 
	}
	\label{fig:band_structure}
	\end{center}
\end{figure}

Recently, high-order EPs have attracted increasing interest due to their diversity of topological features and sophisticated energy band structure beyond EP2s~\cite{HPEn_Feature1}  and have been experimentally observed in classical systems~\cite{Nexus1, Brids2, ECircuits2019, ECircuits2023} and quantum systems~\cite{Wang2023EP3, WuyangNVEP3, HuyingatomEP3,entangleforEP}. In non-Hermitian systems with more than two states, multiple EPs will interact and their coalescence will produce high-order EPs, resulting in novel phenomena that reflect new physics. Generally, an $n$th ($n\ge 2)$ order EP (EPn) can be obtained in a $2n-2$ dimensional parameter space, where $n$ eigenvalues and eigen-states coalesce~\cite{EP_para_demand}. Exquisite control has been developed in a few experimental systems to achieve EPn's,
such as the dilation method using an ancilla qubit in the nitrogen-vacancy system to demonstrate a third-order exceptional line~\cite{WuyangNVEP3}, the combination of a series of operations in the photon interferometry to observe topologically stable third-order EPs and simulate fourfold degeneracies~\cite{Wang2023EP3}, and the collective evolution of the atomic ensemble with a single tunable dissipative parameter to realize an EP nexus from the coalescence of two exceptional arcs with distinct geometries~\cite{HuyingatomEP3}. 
Recently, trapped ion systems have gained increasing interest for quantum simulation of non-Hermitian physics up to EP2 \cite{Trapped_ions1, Trapped_ions2, Trapped_ions3, Trapped_ions4, Trapped_ions5}. However, experimental demonstration for EPn remains elusive in trapped ion systems, and effort has been made in a recent proposal utilizing a two-qubit system with Ising interaction~\cite{theoryEP3_twoions}.

Here, we develop programmable dissipative and coherent controls for multiple levels in a single trapped $^{40}\rm{Ca}^+$ ion, and demonstrate the dynamics of non-Hermitian Hamiltonians with up to EP4, with 6 independent parameters. 
We simultaneously apply individual-controlled radio-frequency and laser fields to drive multiple energy levels, corresponding to the coherent and dissipative components, natively comprising the target non-Hermitian Hamiltonian. By analyzing the resulting dynamics, we can extract the eigen-energies and observe the EP4 at the degeneracies. Additionally, we showcase the programmable capabilities in high-dimensional parameter space and realize a series of non-Hermitian Hamiltonians. 
We tune parameters along the EP2 space, and observe the coalescence of two EP2's into the EP4. Our experiments extend existing coherent control methods \cite{PRL2013Sherman,Ringbauer2022, Yuan2022DD, Zhang2022topo} to include programmable dissipation, and may pave the way for scalable quantum simulation of high dimensional dissipative systems in trapped ion systems. Our demonstration can also be beneficial for the exploration of topological nature around high-order EPs or other interesting EP geometries.

The general form of non-Hermitian Hamiltonian we are interested in is 
\begin{equation}
    H=
    \begin{pmatrix}
       i\gamma_{1} & J_1/\sqrt{3} & 0 & 0 \\
       J_1/\sqrt{3} & i\gamma_{2} & 2J_2/3 & 0 \\
       0 & 2J_2/3 & i\gamma_{3} & J_3/\sqrt{3} \\
       0 & 0 & J_3/\sqrt{3} & i\gamma_{4}
    \end{pmatrix},
    \label{eq:Hamltonian}
\end{equation}
where diagonal elements are the unitless dissipation rates, and the off-diagonal elements are the unitless Rabi rates of the coherent drives. We define $\bm{J}=(J_1,J_2,J_3)$ and $\bm{\gamma}=(\gamma_{1},\gamma_{2},\gamma_{3},\gamma_{4})$, which need to be individually controlled to realize programmable non-Hermitian Hamiltonian. To illustrate EP4, we consider the case when $\bm{\gamma}=\gamma(1,1/3,-1/3,-1)$ and $\bm{J}=J(1,1,1)$, so the non-Hermitian Hamiltonian of interest can be denoted as
\begin{align}
    H_{4}=g(J\mathbf{X}_4+i\gamma \mathbf{Z}_4),
        \label{eq:HamltonianPT}
\end{align}
where $g$ is the strength factor, $\mathbf{X}_4$ and $\mathbf{Z}_4$ are the normalized spin operator with a given quantum number $\mathbf{S}=3/2$ in $2\mathbf{S}+1$ dimension space~(see  \cite{Q_mechanics_JJS} and supplemental information \cite{SM}). We plot the normalized band structure of $H_4/g$ in Fig.~\ref{fig:band_structure}(a) and (b). The regime with $\gamma<J$ $(\gamma>J)$ gives purely real (imaginary) eigen-energies and the band structure is symmetric to the $\gamma-J$ plane at $\rm{Re(E)=0}$ $(\rm{Im(E)=0})$. The EP4s could be found when $\gamma=J$ where all eigenvalues coalesce. For demonstration,  we focus on the response of the eigen-values to variation of $\gamma$ given $J=1$ fixed, as shown in Fig.~\ref{fig:band_structure}(a) and (b).
 
Furthermore, based on Hamiltonian $H$, we can observe the coalescence of EP2s into EP4 via tracking the relative coherent strengths and dissipation rate in the parameter space. We consider non-Hermitian Hamiltonian similar to Eq.~\ref{eq:HamltonianPT} except a tunable $\bm{J}=(J_1,J_2,J_1)$, which leads to surfaces composed of EP2 in the parameter space and EP4 at the surface endpoints $\{\gamma,J_1,J_2\}=\{1,1,1\}$, as shown in Fig.~\ref{fig:band_structure}(c). We demonstrate two types of coalescence behaviors. Firstly, we fix $\gamma=1$ as an intersecting plane with the EP2 surface, giving EP2 lines depicted as solid lines in Fig.~\ref{fig:band_structure}(c) who merge into EP4 at the point. Secondly, we project the EP2 lines along the direction of $\gamma$ to find another pair of EP2 lines, depicted as dashed lines in  Fig.~\ref{fig:band_structure}(c), where they also merge into EP4 at the end point. The parameters $J_1$ and $J_2$ are mutual for the relevant solid and dashed lines, thus in the experiment we select these $\bm{J}$ parameters, and scan $\gamma$ to reveal the EP2's and their coalescence.

\begin{figure}[t!]
	\begin{center}

	\includegraphics[width=1\columnwidth]{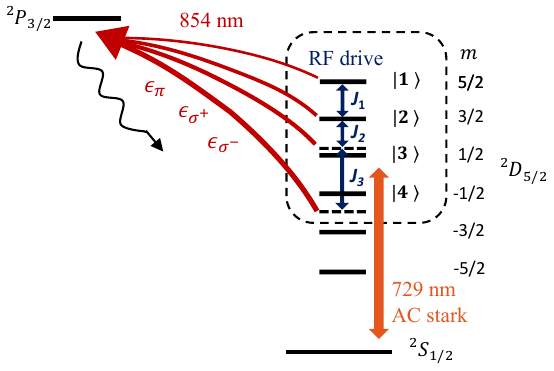}
	\caption{The energy level structure of $^{40}\rm{Ca}^+$ ion and implementation of coherent 
 and dissipative control with trapped ion. A far detuned 729~nm laser is set to generate AC stark between $\ket{3}$ and $\ket{4}$ to distinguish the transitions between adjacent sublevels in $^2D_{5/2}$. Controllable coherent control is realized by radiofrequency drive. 854~nm laser with controlled laser intensity and polarization can pump out of the system to auxiliary levels in the $^2P_{3/2}$ manifold, where the population will basically spontaneously radiate to the ground state $^2S_{1/2}$ soon.   
	}
	\label{fig:Ca_energy}
	\end{center}
\end{figure} 

The experimental setup employs a laser-controlled trapped $^{40}\rm{Ca}^+$ ion in ambient magnetic field at about 0.54~$\rm{mT}$ with multiple energy levels, as illustrated in Fig.~\ref{fig:Ca_energy}. The four-dimensional system in Eq.~\ref{eq:Hamltonian} is constructed with Zeeman sublevels of the $^2D_{5/2}$ manifold with quantum number $m=\{5/2, 3/2, 1/2, -1/2\}$, denoted as $\ket{1}$, $\ket{2}$, $\ket{3}$ and $\ket{4}$, respectively. The experimental methods for implementing coherent and dissipative controls are detailed below.

For coherent control, adjacent sublevels can be coupled by utilizing radiofrequency (RF) fields. However, these transitions are naturally degenerate under the strength of magnetic field mentioned above. To overcome this, we utilize level-selected energy shifts to break the degeneracy, enabling individually controlled coherent operations. We introduce a far-detuned 729~nm laser off-resonantly coupling $\ket{3}$ and $\ket{4}$ states to $^2S_{1/2}$ levels for AC Stark shifts, resulting the three frequency-addressable RF resonant transitions of $9.0976~\rm{MHz}$, $9.0792~\rm{MHz}$, and $9.1146~\rm{MHz}$. We then simultaneously apply all these frequency components using a power-splitter, driving all adjacent transitions resonantly with individually controlled Rabi frequencies. Thus, we achieve the desired coherent components of the Hamiltonian.

\begin{figure}[t!]
	\begin{center}

	\includegraphics[width=1\columnwidth]{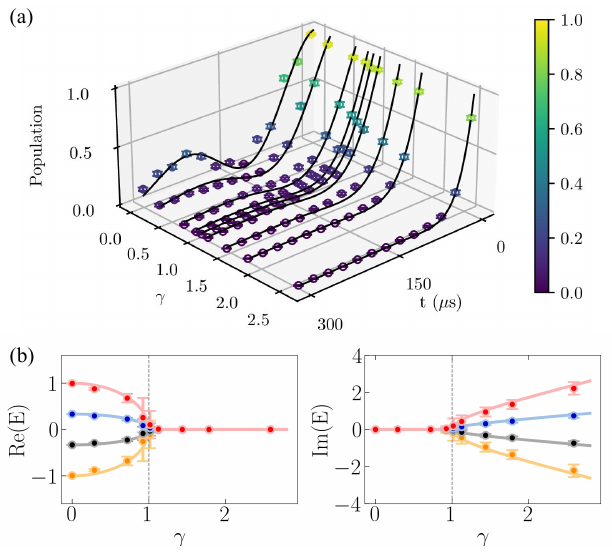}
	\caption{Experimental observation of the EP4. (a) Evolution of the population of $\ket{2}$ state  with controlled parameter $\gamma$. The dots depict the data points, while the black lines represent the theoretical predictions. (b) Real and imaginary part of the eigenvalues of Hamiltonian $H_{4}$. Red, blue, black, and yellow points are the experimental results belonging to different energy bands, and the corresponding solid lines are the theoretical results. While in $\gamma<1(\gamma>1)$ regime only exist pure real(imaginary) eigenvalues, indicating the $\mathcal{PT}$-symmetry unbroken(broken) zone.}
	\label{fig:ep3_eng}
	\end{center}
 
\end{figure}
For dissipative control, an 854~nm laser field with tunable intensity and polarization pumps the populations out of the system into auxiliary levels in the $^2P_{3/2}$ manifold, which then decay rapidly to the ground state $^2S_{1/2}$ via spontaneous emission, as shown in Fig.~\ref{fig:Ca_energy}. The decay rates are individually controlled by adjusting the laser polarization, defined by the vector $\bm{\epsilon}$=$(\epsilon_{\sigma^{+}}, \epsilon_{\sigma^{-}}, \epsilon_{\pi})$, where $\sigma^\pm$ and $\pi$ represent circular and linear components with their summation equal to 1. The pump rates from state $\ket{i}$ are proportional to $\sum_{j=\{\sigma^\pm,\pi\}} |C_{ij}|^2\epsilon_j$, where $C_{ij}$ denotes the Clebsch-Gordan coefficients for the dipole transition matrix elements from $\ket{i}$ to the sublevel of $^2P_{3/2}$~\cite{SM}. 

Adding a global dissipation term proportional to $-i\alpha g\gamma\mathbf{I}$ to the non-Hermitian Hamiltonian does not affect the EP properties \cite{Trapped_ions2, Trapped_ions4}, where $\mathbf{I}$ is the identity matrix, and $\alpha$ is a unit-less parameter. Thus, complete dissipative control of  the four-dimensional system would require three independent parameters, seemingly satisfied by tuning the laser intensities for each polarization components. However, in our experiment, the dipole matrix elements \cite{SM}  enforce a constraint $\gamma_1-3\gamma_2+3\gamma_3-\gamma_4=0$, which coincidentally matches our requirements despite creating a limitation for other purposes. Nevertheless, general control could be achieved by applying a higher magnetic field to enable frequency addressing of the transitions, thereby introducing additional degrees of freedom and offering potential scalability to larger systems. Here for demonstration, we adopt the simplified condition and identify a solution with $\epsilon_{\sigma^{+}}=2/3$, $\epsilon_{\sigma^{-}}=0$, $\epsilon_{\pi}=1/3$ and $\alpha=1$ so that the desired dissipation rates of $-g(\bm{\gamma}-\alpha \mathbf{I})=2g\gamma/3(0,1,2,3)$. Experimentally, the 854~nm laser is split into two paths to provide polarization components of $\sigma^{+}$  and $\pi$, each with tunable intensity. We obtain typical dissipation rates of $0.4(3)~\rm{kHz}, 10.0(4)~\rm{kHz}, 20.4(1.7)~\rm{kHz}$ and $30.3(1.8)~\rm{kHz}$, close to the desired ratio. Then we can vary the laser strengths for desired values of $\gamma$.

\begin{figure*}[ht!]
	\begin{center}
     
	\includegraphics[width=1\columnwidth]{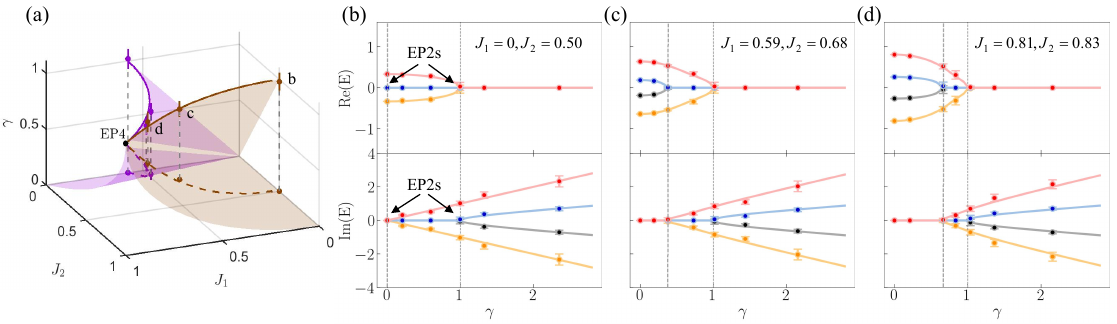}
        \captionsetup{width=2\columnwidth}
	\caption{Experimental observation of the coalescence of EP2s into EP4. (a) The experimental results of the coalescence of EP2s into EP4. The brown dots with error bars are the experimental results for one type of coalescence, with the brown lines the corresponding theoretical prediction. The purple set of data and lines shows another type of coalescence. (b)-(d) Real and imaginary components of the eigenvalues of Hamiltonian which $\{J_{1},J_{2}\}=\{0,0.50\},\{0.59,0.68\},\{0.81,0.83\}$ for the brown solid and dashed line in (a) as examples. Red, blue, black, and yellow points are the experimental results belonging to different energy bands, and the corresponding solid lines are the theoretical predictions. Vertical dashed lines show the identified EP2 points.}
	\label{fig:ep4_eng}
	\end{center}
\end{figure*}

To observe EP4, the energies of the Non-Hermitian system are be extracted by monitoring the system's dynamics with varied $\gamma$. The ion is first Doppler and sideband cooled, then initialized to $\ket{2}$ by optical pumping and resonant laser transition. 
We construct the corresponding Hamiltonian by applying the detuned 729~nm laser, RF fields, and 854~nm laser fields. We set $g=2\pi\times2.3~\rm{kHz}$ by tuning  strengths of the coherent and dissipation, and set the values of $\bm{\gamma}$ and $\bm{J}$ as mentioned above. Afterwards, we measure the populations in $\ket{2}$. 
Figure~\ref{fig:ep3_eng}(a) depicts the experimental results of the evolution of $P_{\ket{2}}$ with controlled parameters $\gamma$ and the black lines are corresponding theoretical curves. $P_{\ket{2}}$ oscillates  at $\gamma=0.00$ where the strength of gain is larger than that of loss but behaves like exponential decay at $\gamma=2.59$ where loss dominates the evolution. To extract eigen-energies of $H_4$ and observe the EP4, we adopt the technique in \cite{Trapped_ions2, Trapped_ions1} and apply numerical curve fitting to the measured $P_{\ket{2}}$ data with the theoretical population under $H_{4}-i\alpha g\gamma \mathcal{I}$, where we treat $\gamma$ as the only free parameter. We then diagonalize the extracted Hamiltonian $H_4^\prime$ and obtain the eigen-energies normalized to $g$.  Figure~\ref{fig:ep3_eng}(b) illustrate the eigen-energies of $H_4^\prime$ with varied $\gamma$, pure real (imaginary) values in regime with $\gamma<1$ $(\gamma>1)$ and coalesce at $\gamma=1$ where EP4 exists. The transition of eigen-energies from purely real to purely imaginary also indicates the $\mathcal{PT}$-symmetry broken transition.

To observe the coalescence of EP2s into EP4, we maintain the settings of the dissipation components while adjust the strengths of the coherent drives, as mentioned above. 
By setting coherent driving strength for selected points along the EP2 lines as in Fig.~\ref{fig:band_structure} and scanning the parameter $\gamma$, 
we experimentally locate EP2s and observe their coalescence, as shown in Fig.~\ref{fig:ep4_eng}(a) and detailed below. For the brown solid and dashed lines in Fig.~\ref{fig:ep4_eng}(a), we choose three points corresponding to $\{J_{1},J_{2}\}=\{0,0.50\},\{0.59,0.68\},\{0.81,0.83\}$. Fig.~\ref{fig:ep4_eng}(b)-(d) illustrate the eigen-energies of these three positions with varied $\gamma$. In Fig.~\ref{fig:ep4_eng}(b), there are only three different eigen-energies when $\gamma=0$ and when $\gamma=1$, 
which indicates that two of the eigen-energies degenerate. Since Fig.~\ref{fig:ep4_eng}(c) shows EP2s where the blue and black lines merge, we can indicate the degenerate energies for the EP2s shown at Fig.~\ref{fig:ep4_eng}(b).
As $J_1$ and $J_2$ get closer to EP4 in Fig.~\ref{fig:ep4_eng}(c), we observe the two EP2s get closer to each other, eventually coalesce into EP4 in Fig.~\ref{fig:ep3_eng}(b). For the purple lines in Fig.~\ref{fig:ep4_eng}(a), we choose another three points which $\{J_{1},J_{2}\}=\{0.58,0\},\{0.71,0.60\},\{0.81,0.77\}$ and repeat the above measurement, and we observe similar coalescence behavior as detailed in \cite{SM}.

In conclusion, we present a programmable multi-level control technique using a trapped-ion system to experimentally investigate the band structure of a non-Hermitian Hamiltonian in a high-dimensional parameter space. By leveraging controllable dissipation and coherent drives, we demonstrate the existence of EP4 and the coalescence of multiple EP2s into EP4. This work highlights the experimental feasibility of exploring high-dimensional non-Hermitian systems, paving the way for further investigations of novel phenomena within trapped ion systems. Our approach is versatile and can be extended to other platforms, such as superconducting circuits, quantum dots, and atom arrays controlled by optical tweezers. Our results can also be applied to the exploration of topological features of the complex energy surface around high-order EPs \cite{Brids2,Nexus1} and various EP geometries, such as lines or rings formed entirely by high-order EPs \cite{Nat_Rev_Phys}. Moreover, combining this demonstration with ion-ion coupling in scaled ion chains or 2D ion crystals may provide a complex programmable platform to explore applications such as many-body open quantum systems and phase transitions. Besides our demonstration, we noticed during the preparation of this work, that a third-order exceptional point was experimentally demonstrated in a dissipative trapped ion system \cite{Duan3}.

\begin{acknowledgments}
We thank Kamran Rehan and Waner Hou for their support on the experimental setup and Wei Zhang and Xiang Zhang for their helpful comments on the manuscript. We acknowledge support from the National Natural Science Foundation of China (grant number 12261160569, 92165206, 12174373
), Innovation Program for Quantum Science and Technology (Grant No. 2021ZD0301603). 
\end{acknowledgments}

\newpage

\section{Supplemental Material}

\subsection{Theoretical Details}
\emph{Normalized high dimensional spin operator.---}For second-order situations, the normalized spin operators are the Pauli operators. We label the normalized spin operators in fourth-order situation as $\mathbf{F}_4$, where $\mathbf{F}=\{\mathbf{X}, \mathbf{Z}\}$, thus
\begin{equation}
    \mathbf{X}_4=\left(
    \begin{matrix}
         0 & 1/\sqrt{3} & 0 & 0\\
         1/\sqrt{3} & 0 & 2/3 & 0\\
         0 & 2/3 & 0 & 1/\sqrt{3} \\
         0 & 0 & 1/\sqrt{3} & 0
    \end{matrix}\right)
\end{equation}

\begin{equation}
    \mathbf{Z}_4=\left(
    \begin{matrix}
         1 & 0 & 0 & 0\\
         0 & 1/3 & 0 & 0\\
         0 & 0 & -1/3 & 0\\
         0 & 0 & 0 & -1
    \end{matrix}\right)
\end{equation}

\subsection{Experimental Details}
\begin{figure}[t!]
	\begin{center}

	\includegraphics[width=0.95\columnwidth]{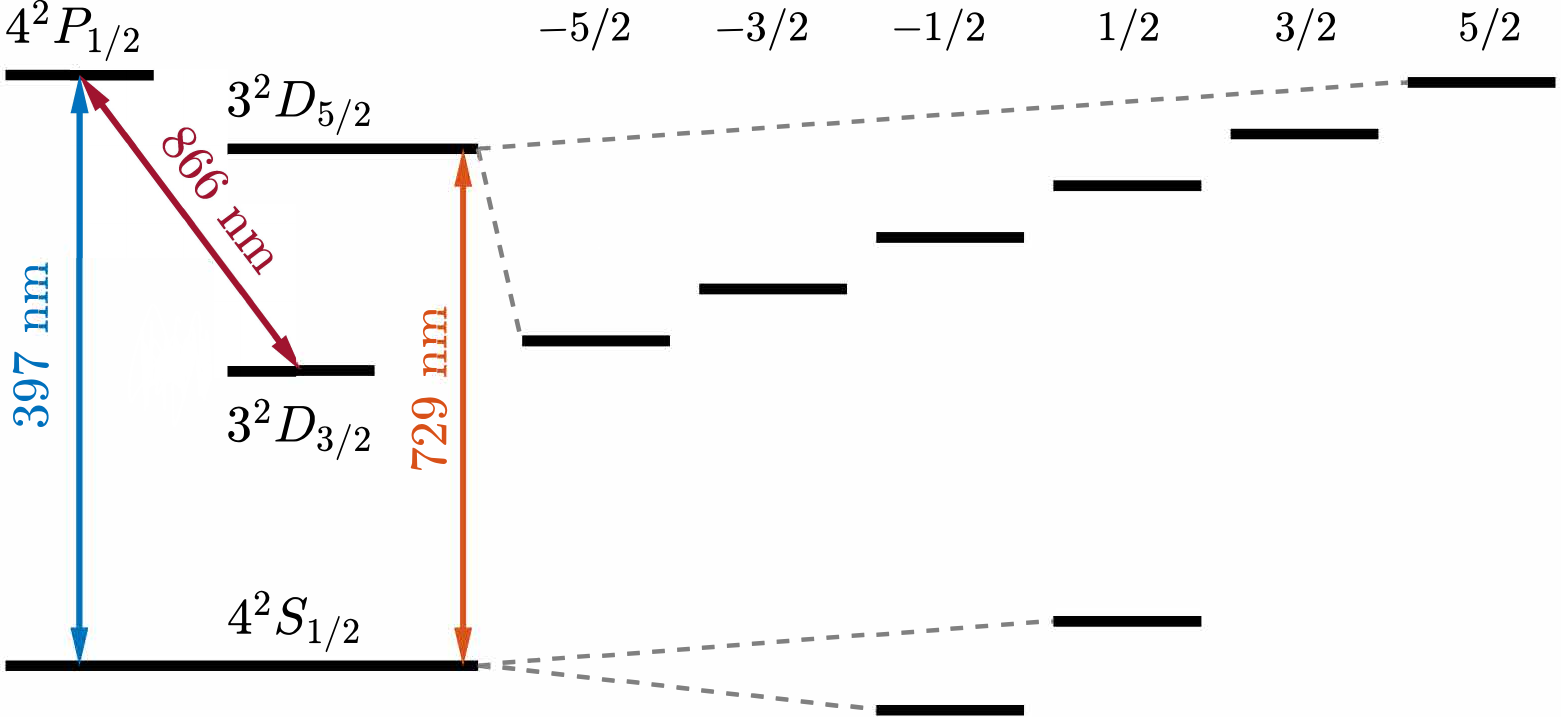}
	\caption{Detail energy level structure of $^{40}\rm{Ca}^+$ ion. The solid arrows denote the laser involved in the shelving and detection. 729~nm laser is used to transfer the states in $^2S_{1/2}$ manifold into $^2D_{5/2}$ manifold. 397~nm and 866~nm laser are used to detect the population via cycling transition.
	}
	\label{fig:Ca_eng_level}
	\end{center}
\end{figure}

\emph{Shelving and Detection.---} For $^{40}\rm{Ca}^+$ ion, the fine structure of $^2S_{1/2}$ and $^2D_{5/2}$ manifold can be described as total angular momentum $J$ and $m_J$ as illustrated in Fig.~\ref{fig:Ca_eng_level}. We use the 397~nm laser to achieve resonance transition and obtain fluorescence counting to detect the population of the states in $^2S_{1/2}$ manifold. To detect the population of $\ket{2}$, we need shelve all the population to $^2S_{1/2}$. However, during the experimental evolution process, the population pumped away by the 854~nm laser in $^2D_{5/2}$ will fall onto $^2S_{1/2}$, so we need to transfer the states in $^2S_{1/2}$ manifold into $^2D_{5/2}$ manifold first. Taking observation of EP4 as an example, after the evolution sequence, we first transfer $\ket{S_{1/2},m_J=-1/2}$ and $\ket{S_{1/2},m_J=1/2}$ to $\ket{D_{5/2},m_J=-5/2}$ and $\ket{D_{5/2},m_J=-3/2}$ with two 729~nm $\pi$ pulses. Then we shelve all the population in $\ket{2}$ to $\ket{S_{1/2},m_J=1/2}$ with a 729~nm $\pi$ pulses. After shelving, we can detect the corresponding population via cycling transition driven by 397~nm and 866~nm laser.

%
\begin{figure}[t!]
	\begin{center}

	\includegraphics[width=1\columnwidth]{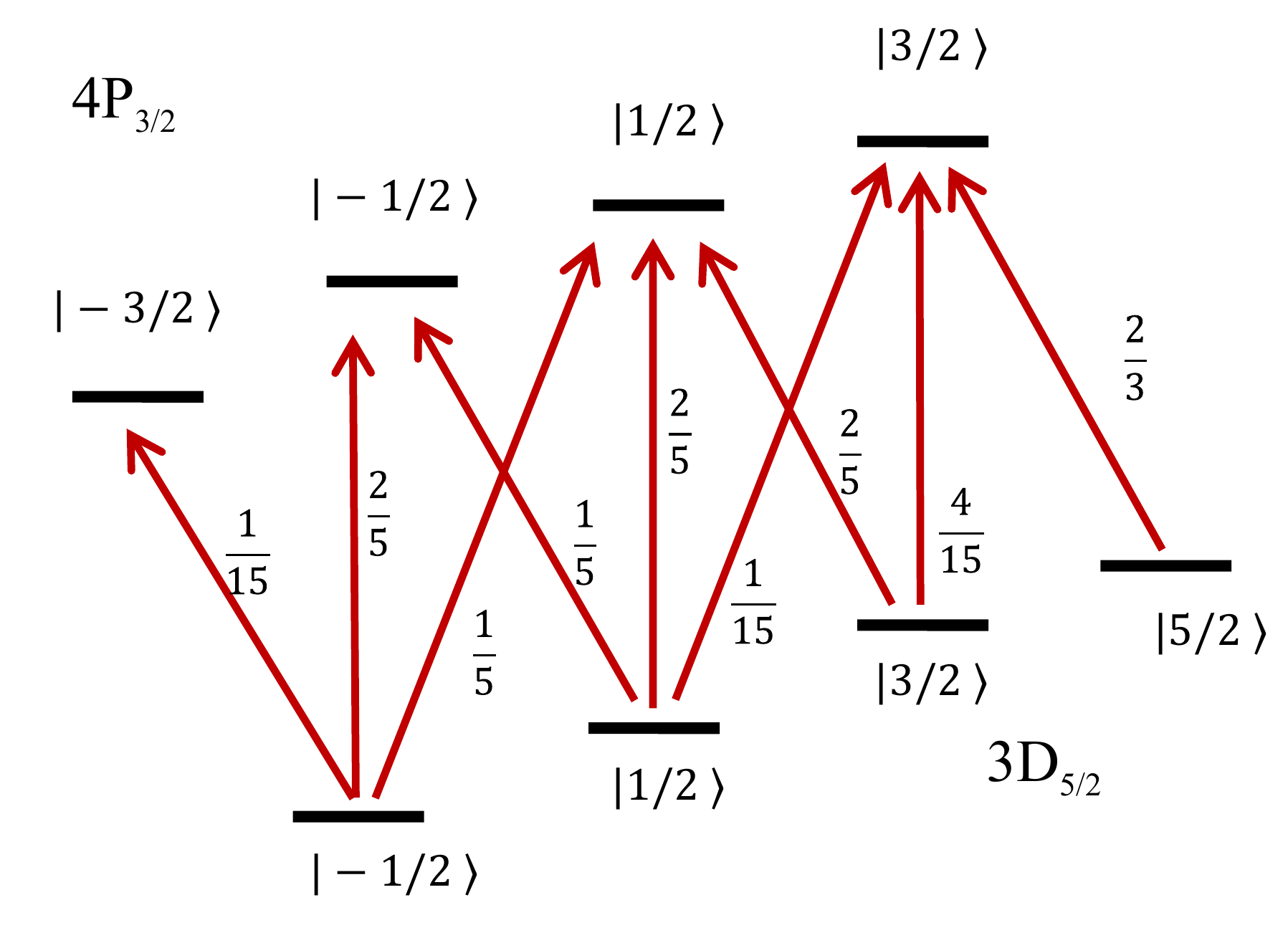}
	\caption{The square of Clebsch-Gordan coefficients for the transition from the $D_{5/2}$ manifold to the $P_{3/2}$ manifold. Due to the fact that the linewidth of 854~nm can basically cover the transitions between $D_{5/2}$ and $P_{3/2}$, so we use these matrix elements to calculate appropriate polarization of 854~nm laser for dissipation.
	}
	\label{fig:trotter_order}
	\end{center}
\end{figure}

\begin{figure*}[t!]
	\begin{center}

	\includegraphics[width=1\columnwidth]{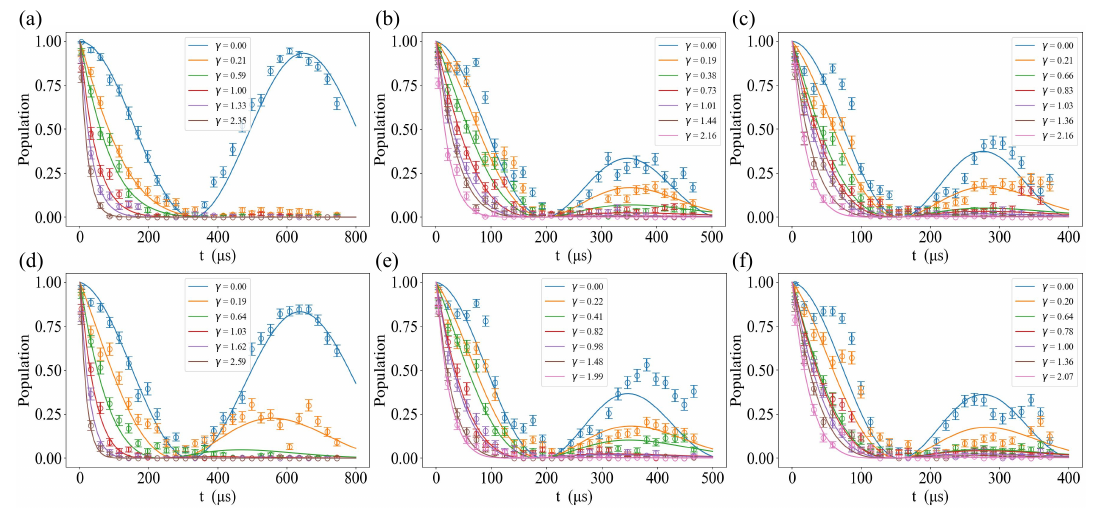}
	\caption{Population evolution corresponding to the coalescence of EP2s into EP4.(a)(b)(c) represent the population evolution of Fig. 4(a)(b)(c). (d)(e)(f) represent the population evolution of Fig.~\ref{fig:J2_EP2toEP4}(a)(b)(c).The solid lines describe the ideal dynamical evolution and the markers represent the experimental results at the same dissipation and coupling strength as that in the ideal dynamical simulation. 
	}
	\label{fig:trotter_order_evo}
	\end{center}
\end{figure*}

\begin{figure*}[t!]
	\begin{center}

	\includegraphics[width=1\columnwidth]{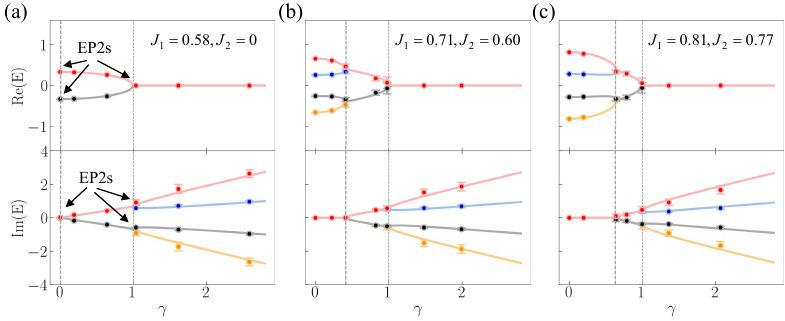}
	\caption{Experimental observation of the coalescence of EP2s into EP4. (a)(b)(c) Real and imaginary part of the eigenvalues of Hamiltonian 
 for three points which $J_{2}=0,0.60,0.77$ on the yellow line in Fig. 4(d). orange circles, black inverted triangles, blue regular triangles, and red stars are the experimental results belonging to different energy bands, the corresponding solid lines are the theoretical results. 
	}
	\label{fig:J2_EP2toEP4}
	\end{center}
\end{figure*}
\emph{Generation of AC Stark Shifts.---} In order to distinguish the transitions between adjacent sublevels in $^2D_{5/2}$ which we care, we apply a detuned 729~nm laser to generate AC Stark shifts. We set the frequency of the laser approximately 3~\rm{MHz} red detuned relative to the transition frequency between $\ket{S_{1/2},m_J=1/2}$ 
and $\ket{D_{5/2},m_J=1/2}$. By fine-tuning the frequency of the laser, we can distinguish these frequencies well.

\emph{Energy extraction and additional data---}
Thanks to the precise calibration of coupling strength $g$, coherent strengths $\bm J$, and the relative dissipation ratios, the target Hamiltonian model mentioned in the main text is utilized for curve fitting populations of $\ket{2}$ with $\gamma$ as the only free parameter. Thus, we obtain the fitted $\gamma^\prime$, leading to an extracted Hamiltonian, whose diagonalization gives the desired energy of the system $E$. We estimate the errorbar via nonparametric bootstrapping method \cite{Bootstrap,nonpara_Bootstrap}, assuming that the raw data follow a normal distribution, and get samples 200 times from the raw data. In each sample, we can fit the data and obtain a series of $\gamma_{s}$, leading to sampled energies $E_{s}$. The error bar of $E$ is obtained by analyzing the statistics of $E_{s}$ array with a 68\% confidence interval. 
Thus we obtain the data with errorbar shown in the main text. In addition, we apply the same technique to observe the coalescence of EP2s into EP4, with the data shown in Fig.~\ref{fig:trotter_order_evo} with the analyzed result shown in Fig.~\ref{fig:J2_EP2toEP4}.


\begin{thebibliography}{100}

\bibitem{Unidirect1} Z. Lin, H. Ramezani, T. Eichelkraut, T. Kottos, H. Cao, and D. N. Christodoulides,~\href{https://link.aps.org/doi/10.1103/PhysRevLett.106.213901}{Phys. Rev. Lett. \textbf{106}, 213901 (2011)}

\bibitem{Unidirect2} A. Regensburger, C. Bersch, M.-A. Miri, G. Onishchukov, D. N. Christodoulides, and U. Peschel,~\href{https://www.nature.com/articles/nature11298}{Nature \textbf{488}, 167–171 (2012)}

\bibitem{Unidirect3} L. Feng, Y.-L. Xu, W. S. Fegadolli, M.-H. Lu, J. E. B. Oliveira, V. R. Almeida, Y.-F. Chen, and A. Scherer,~\href{https://www.nature.com/articles/nmat3495}{Nature Mater \textbf{12}, 108–113 (2013)}

\bibitem{Unidirect4} A. Guo, G. J. Salamo, D. Duchesne, R. Morandotti, M. Volatier-Ravat, V. Aimez, G. A. Siviloglou, and D. N. Christodoulides,~\href{https://link.aps.org/doi/10.1103/PhysRevLett.103.093902}{Phys. Rev. Lett. \textbf{103}, 093902 (2009)}

\bibitem{Single_mode1} L. Feng, Z. J. Wong, R.-M. Ma, Y. Wang, and X. Zhang,~\href{https://www.science.org/doi/10.1126/science.1258479}{Science \textbf{346}, 972–975 (2014)}


\bibitem{Power_transfer1} S. Assawaworrarit, X. Yu, and S. Fan,~\href{https://www.nature.com/articles/nature22404}{Nature \textbf{546}, 387–390 (2017)}


\bibitem{NA_sensing} Z. Li, C. Li, Z. Xiong, G. Xu, Y. R. Wang, X. Tian, X. Yang, Z. Liu, Q. Zeng, R. Lin, Y. Li, J. K. W. Lee, J. S. Ho, and C.-W. Qiu,~\href{https://link.aps.org/doi/10.1103/PhysRevLett.130.227201}{Phys. Rev. Lett. \textbf{130}, 227201 (2023)}


\bibitem{HeatTransport2023} G. Xu, X. Zhou, Y. Li, Q. Cao, W. Chen, Y. Xiao, L. Yang, and C.-W. Qiu,~\href{https://link.aps.org/doi/10.1103/PhysRevLett.130.266303}{Phys. Rev. Lett. \textbf{130}, 266303 (2023)}

\bibitem{RMP_EPtopology} Emil J. Bergholtz, Jan Carl Budich, and Flore K. Kunst,~\href{https://link.aps.org/doi/10.1103/RevModPhys.93.015005}{Rev. Mod. Phys. \textbf{93}, 015005 (2021)}



\bibitem{Fermic_arc1} H. Zhou, C. Peng, Y. Yoon, C. W. Hsu, K. A. Nelson, L. Fu, J. D. Joannopoulos, M. Solja{\v{c}}i{\'{c}}, and B. Zhen,~\href{https://www.science.org/doi/10.1126/science.aap9859}{Science \textbf{359}, 1009–1012 (2018)}.

\bibitem{Wely_ring1} A. Cerjan, S. Huang, M. Wang, K. P. Chen, Y. Chong, and M. C. Rechtsman,~\href{https://www.nature.com/articles/s41566-019-0453-z}{Nat. Photonics \textbf{13}, 623–628 (2019)}

\bibitem{Brids1} H. Hu, S. Sun, and S. Chen,~\href{https://link.aps.org/doi/10.1103/PhysRevResearch.4.L022064}{Phys. Rev. Research \textbf{4}, L022064 (2022)}

\bibitem{encircle_EP1} J. Doppler, A. A. Mailybaev, J. B{\"o}hm, U. Kuhl, A. Girschik, F. Libisch, T. J. Milburn, P. Rabl, N. Moiseyev, and S. Rotter,~\href{https://www.nature.com/articles/nature18605}{Nature \textbf{537}, 76–79 (2016)}

\bibitem{encircle_EP2} Q. Zhong, M. Khajavikhan, D. N. Christodoulides, and R. El-Ganainy,~\href{https://www.nature.com/articles/s41467-018-07105-0}{Nat Commun \textbf{9}, 4808 (2018)}

\bibitem{encircle_EP3} X.-L. Zhang, S. Wang, B. Hou, and C. T. Chan,~\href{https://link.aps.org/doi/10.1103/PhysRevX.8.021066}{Phys. Rev. X \textbf{8}, 021066 (2018)}

\bibitem{encircle_EP4} F. Yu, X.-L. Zhang, Z.-N. Tian, Q.-D. Chen, and H.-B. Sun,~\href{https://link.aps.org/doi/10.1103/PhysRevLett.127.253901}{Phys. Rev. Lett. \textbf{127}, 253901 (2021)}

\bibitem{NV_PT1} Y. Wu, W. Liu, J. Geng, X. Song, X. Ye, C.-K. Duan, X. Rong, and J. Du
,~\href{https://www.science.org/doi/10.1126/science.aaw8205}{Science \textbf{364}, 878–880 (2019)}

\bibitem{Electronic_oscillator1} N. Bender, S. Factor, J. D. Bodyfelt, H. Ramezani, D. N. Christodoulides, F. M. Ellis, and T. Kottos,~\href{https://link.aps.org/doi/10.1103/PhysRevLett.110.234101}{Phys. Rev. Lett. \textbf{110}, 234101 (2013)}

\bibitem{Ultracold_atoms1} J. Li, A. K. Harter, J. Liu, L. de Melo, Y. N. Joglekar, and L. Luo,~\href{https://www.nature.com/articles/s41467-019-08596-1}{Nat Commun \textbf{10}, 855 (2019)}

\bibitem{Ultracold_atoms2} Z. Ren, D. Liu, E. Zhao, C. He, K. K. Pak, J. Li, and G.-B. Jo,~\href{https://www.nature.com/articles/s41567-021-01491-x}{Nat. Phys. \textbf{18}, 385–389 (2022)}

\bibitem{Thermal_atoms} C. Liang, Y. Tang, A.-N. Xu, and Y.-C. Liu,~\href{https://link.aps.org/doi/10.1103/PhysRevLett.130.263601}{Phys. Rev. Lett. \textbf{130}, 263601}


\bibitem{Trapped_ions2} L. Ding, K. Shi, Q. Zhang, D. Shen, X. Zhang, and W. Zhang,~\href{https://link.aps.org/doi/10.1103/PhysRevLett.126.083604}{Phys. Rev. Lett. \textbf{126}, 083604 (2021)}

\bibitem{Trapped_ions1} L. Ding, K. Shi, Y. Wang, Q. Zhang, C. Zhu, L. Zhang, J. Yi, S. Zhang, X. Zhang, and W. Zhang,~\href{https://link.aps.org/doi/10.1103/PhysRevA.105.L010204}{Phys. Rev. A \textbf{105}, L010204 (2022)} 

\bibitem{Trapped_ions3} L.-L. Yan, J.-W. Zhang, M.-R. Yun, J.-C. Li, G.-Y. Ding, J.-F. Wei, J.-T. Bu, B. Wang, L. Chen, S.-L. Su, F. Zhou, Y. Jia, E.-J. Liang, and M. Feng,~\href{https://link.aps.org/doi/10.1103/PhysRevLett.128.050603}{Phys. Rev. Lett. \textbf{128}, 050603 (2022)}

\bibitem{Trapped_ions4} W.-C. Wang, Y.-L. Zhou, H.-L. Zhang, J. Zhang, M.-C. Zhang, Y. Xie, C.-W. Wu, T. Chen, B.-Q. Ou, W. Wu, H. Jing, and P.-X. Chen,~\href{https://link.aps.org/doi/10.1103/PhysRevA.103.L020201}{Phys. Rev. A \textbf{103}, L020201 (2021)}

\bibitem{Trapped_ions5} M.-M. Cao, K. Li, W.-D. Zhao, W.-X. Guo, B.-X. Qi, X.-Y. Chang, Z.-C. Zhou, Y. Xu, and L.-M. Duan,~\href{https://link.aps.org/doi/10.1103/PhysRevLett.130.163001}{Phys. Rev. Lett. \textbf{130}, 163001 (2023)}

\bibitem{HPEn_Feature1} I. Mandal and E. J. Bergholtz,~\href{https://link.aps.org/doi/10.1103/PhysRevLett.127.186601}{Phys. Rev. Lett. \textbf{127}, 186601 (2021)}

\bibitem{Nexus1} W. Tang, X. Jiang, K. Ding, Y.-X. Xiao, Z.-Q. Zhang, C. T. Chan, and G. Ma,~\href{https://www.science.org/doi/10.1126/science.abd8872}{Science \textbf{370}, 1077–1080 (2020)}

\bibitem{Brids2} Y. S. S. Patil, J. H{\"o}ller, P. A. Henry, C. Guria, Y. Zhang, L. Jiang, N. Kralj, N. Read, and J. G. E. Harris,~\href{https://www.nature.com/articles/s41586-022-04796-w}{Nature \textbf{607}, 271–275 (2022)} 


\bibitem{ECircuits2023} K. Bai, J.-Z. Li, T.-R. Liu, L. Fang, D. Wan, and M. Xiao,~\href{https://link.aps.org/doi/10.1103/PhysRevLett.130.266901}{Phys. Rev. Lett. \textbf{130}, 266901 (2023)}

\bibitem{ECircuits2019} Z. Xiao, H. Li, T. Kottos, and A. Alù,~\href{https://link.aps.org/doi/10.1103/PhysRevLett.123.213901}{Phys. Rev. Lett. \textbf{123}, 213901 (2019)}

\bibitem{Wang2023EP3} K. Wang, L. Xiao, H. Lin, W. Yi, E. J. Bergholtz, and P. Xue,~\href{https://www.science.org/doi/10.1126/sciadv.adi0732}{Sci. Adv. \textbf{9}, eadi0732 (2023).}

\bibitem{WuyangNVEP3} Y. Wu, Y. Wang, X. Ye, W. Liu, Z. Niu, C. K. Duan, Y. Wang, X. Rong, and J. Du,~\href{https://doi.org/10.1038/s41565-023-01583-0}{Nat. Nanotechnol. \textbf{19}, 160–165 (2024)}

\bibitem{HuyingatomEP3} C. Wang, N. Li, J. Xie, C. Ding, Z. Ji, L. Xiao, S. Jia, B. Yan, Y. Hu, and Y. Zhao,~\href{https://doi.org/10.1103/PhysRevLett.132.253401}{Phys. Rev. Lett. \textbf{132}, 253401 (2024)}

\bibitem{entangleforEP} Z. Tang, T. Chen, X. Tang, X. D. Zhang,~\href{https://doi.org/10.1038/s41377-024-01514-1}{Light Sci. Appl. \textbf{13}, 167 (2024)}

\bibitem{EP_para_demand} J. H{\"o}ller, N. Read, and J. G. E. Harris,~\href{https://link.aps.org/doi/10.1103/PhysRevA.102.032216}{Phys. Rev. A \textbf{102}, 032216 (2020)}


\bibitem{theoryEP3_twoions} T. T. Shi,  L. D. Zhang, S. N. Zhang, W. Zhang,~\href{https://wulixb.iphy.ac.cn/article/doi/10.7498/aps.70.20220716}{Acta Phys. Sin. \textbf{71}, 130303 (2022)}

\bibitem{PRL2013Sherman} 
J. A. Sherman, M. J. Curtis, D. J. Szwer, D. T. C. Allcock, G. Imreh, D. M. Lucas, and A. M. Steane, \href{https://doi.org/10.1103/PhysRevLett.111.180501}{
Phys. Rev. Lett. \textbf{111}, 180501 (2013)}


\bibitem{Yuan2022DD} X. Yuan, Y. Li, M. Zhang, C. Liu, M. Zhu, X. Qin, N. V. Vitanov, Y. Lin, and J. Du,~\href{https://link.aps.org/doi/10.1103/PhysRevA.106.022412}{Phys. Rev. A \textbf{106}, 022412 (2022)}

\bibitem{Zhang2022topo} M. Zhang, X. Yuan, Y. Li, X.-W. Luo, C. Liu, M. Zhu, X. Qin, C. Zhang, Y. Lin, and J. Du,~\href{https://link.aps.org/doi/10.1103/PhysRevLett.129.250501}{Phys. Rev. Lett. \textbf{129}, 250501 (2022)}
\bibitem{Ringbauer2022} M. Ringbauer, M. Meth, L. Postler, R. Stricker, R. Blatt, P. Schindler, amd T. Monz,~\href{https://doi.org/10.1038/s41567-022-01658-0}{Nature Physics \textbf{18}, 1053 (2022)}

\bibitem{Q_mechanics_JJS} J. J. Sakurai and Jim Napolitano, \emph{Modern Quantum Mechanics} ~\href{https://doi.org/10.1017/9781108587280}
{(Cambridge University Press,3rd edition, 2020)}

\bibitem{SM} See Supplementary materials for details about the model, the experimental setup, the detecting method, numerical simulation, and the analysis of experimental data.

\bibitem{Bootstrap} Gareth James, Daniela Witten, Trevor Hastie and Robert Tibshirani, \emph{An Introduction to Statistical Learning} (2nd edition)

\bibitem{nonpara_Bootstrap} Anthony Kulesa, Martin Krzywinski, Paul Blainey, and Naomi Altman, \href{https://doi.org/10.1038/nmeth.3414}{Nature Methods \textbf{12}, 477–478 (2015)}


\bibitem{Nat_Rev_Phys} K. Ding, C. Fang, G. Ma, \href{https://doi.org/10.1038/s42254-022-00516-5}{Nat. Rev. Phys. \textbf{4}, 745–760 (2022)}.

\bibitem{Duan3} Y.-Y. Chen, K. Li, L. Zhang, Y.-K. Wu, J.-Y. Ma, H.-X. Yang, C. Zhang, B.-X. Qi, Z.-C. Zhou, P.-Y. Hou, Y. Xu, L.-M. Duan, \href{https://doi.org/10.48550/arXiv.2412.05870}{arXiv:2412.05870 (2024)}. 





\end{thebibliography}
\end{document}